\documentclass[12pt,epsf]{article}
\usepackage[pctex32]{graphics}
\makeatletter

\@addtoreset{equation}{section}
\newcommand{\be}{\begin{eqnarray}}
\newcommand{\ee}{\end{eqnarray}}
\newcommand{\ba}{\begin{array}}
\newcommand{\ea}{\end{array}}
\newcommand{\nn}{\nonumber}

\makeatletter \@addtoreset{equation}{section} \makeatother

\begin{document}
\vspace{1cm}
\begin{center}
~\\~\\~\\~\\~\\
{\bf  \Large  KK6 from M2 in ABJM}
\vspace{2cm}

                      Wung-Hong Huang\\
                       Department of Physics\\
                       National Cheng Kung University\\
                       Tainan, Taiwan\\

\end{center}
\vspace{2cm}
\begin{center}{\bf  \Large ABSTRACT } \end{center}
We study the BPS equations in ABJM theory furthermore.   Besides the known funnel solutions which are proposed to describe M2-M5 systems  we  find a new kind solution.  We analyze the properties therein and see that it can describe M2-KK6 (2D membranes with 6D Kaluza-Klein monopole) systems.   Our solutions also reveal new vacua in the mass deformed ABJM theory.  Contrast to the previous investigation in which the vacua are reducible, ours may be irreducible.  We also discuss a possible interpretation of  the new BPS solutions as a M2-2M5 system.

\vspace{4cm}
\begin{flushleft}
*E-mail:  whhwung@mail.ncku.edu.tw\\
\end{flushleft}
%%%%%%%%%%%%%%%%%%%%%%%
\newpage
\section{Introduction}
Bagger, Lambert and Gustavsson  (BLG model) [1,2] had proposed a three-dimensional N = 8 superconformal Chern-Simons model based a novel 3-Lie algebra [3].  This model has been expected to describe the low energy effective theory of two coincident M2-branes in eleven dimensions [4]. 

The use of the Nambu bracket algebras  [5,6], which are a infinite-dimensional case of n-Lie algebras, in the context of the BLG model had found that it is possible  to have 3 dimensional internal spaces coming from 3-algebra to combine with the 3 dimensional world-volume of M2 branes to describe M5 [7]. 

  However, in M-theory there are Kaluza-Klein monopole (KK6) object which is a six dimensional object with coordinate ($t,x_1,x_2,z_1,z_2,y_1,y_2$).   It was argued in [8-11] that  M2-brane, with coordinate ($t,x,z$), can intersect with  KK6 over a 0-brane such that one of the M-2-brane coincides with the isometry direction $(z)$ of the Taub-NUT space [8]
\\
\be
 \left(0|M2,KK6\right)=\Big\{\ba{lrcrrrrrrrr}
t    & x & z&-&-&-&-&-&-&-&-  \\
t  & - & (z)&x_1&x_2&z_1&z_2&y_1&y_2&- &-
\ea
\ee
\\
Thus,  in considering $(0|M2,KK6)$ we need extra 5 dimensional internal spaces from 3-algebra in M2 brane to have 6 space dimensions in the world volume of KK6. In [12] we had proved  that the infinite dimensional Lie 3-algebra based on the Nambu-Poisson structure could not only provide three dimensional manifolds to allow M5 from M2 [7], but also provide five dimensional manifolds to allow KK6 from M2.    

    After discovery BLG theory, ABJM [13-15] had proposed an N = 6 Chern-Simons-matter theory with $U(N) \times U(N)$ gauge group, $SO(6)$ R-symmetry and equal but opposite Chern-Simons (CS) levels (k,-k), to capture the dynamics of the low-energy limit of multiple M2-branes on an M-theory orbifold, $C^4/Z_k$.  An important test for the multiple membranes theory is that it should reproduce the physics of M2-M5 intersections.  The studies in [16-18] have confirmed the property. 

  However, as mentioned before,  in M-theory there are Kaluza-Klein monopole (KK6) object and we need to produce the funnel solution of  $(0|M2,KK6)$ systems.   This seem to conflict to the conjecture in [17], which said  that the funnel of M2-M5 are the only BPS solution in ABJM theory.   In this paper  we will solve the puzzle by studying the BPS equations in ABJM theory furthermore.   We will find a new kind solution which can describe the M2-KK6 systems.  We also see that it can describe the M2-2M5 system.

Note that ABJM model admits  a mass deformation which preserves  supersymmetry while being noncomformal  [17,19]. The deformed theory is expected to have a discrete set of vacua labeled by partitions of N, where N is the number of M2-branes. These vacua are dual to a discrete set of LLM geometries constructed in [20] and had been counted in mass deformed ABJM in [17, 21].  As a byproduct, our solutions also show new vacua in the mass deformed ABJM theory.  Contrast to the previous investigation [17,21] in which the vacua are reducible, ours may be irreducible.

This paper is organized as follows.  In section 2 we review the ABJM model and present the BPS equation as those in [18].   In section 3 we  present our numerical  searches for BPS solutions in low dimensional matrix.  In section 4 we first provide  the general solution.   Next, we calculate the tension of the solution and show a possible way to explain our solution as  describing the  M2-KK6 (2D membranes with 6D Kaluza-Klein monopole) systems.  In section 5 we mention that our solutions also reveal new vacua in the mass deformed ABJM theory.  It is interesting to see that, contrast to the previous investigation in which the vacua are reducible, ours may be  irreducible.   We also discuss a possible interpretation of  the new BPS solutions as a M2-2M5 system.  We summarize our results in the last section.
%%%%%%%%%%%%%%
\section {ABJM Theory and BPS Solutions}
%%%%%%%%%%%%%%
\subsection {ABJM Theory and BPS Equations}
The ABJM theory is an N = 6 superconformal $U(N)\times U(N)$ Chern-Simons theory of gauge fields $A_\mu$ and $\hat{A}_\mu$ with level (k,-k) coupled to four complex scalars $Y^A$ and four Dirac fermions $\psi_A$, where $A=1,2,3,4$ in the bifundamental representation [13], 
\be 
S=\int d^3 x \left[ \frac{k}{4 \pi} \varepsilon^{\mu\nu\lambda} \mathrm{Tr} \left(A_{\mu} \partial _{\nu} A_\lambda + \frac{2i}{3} A_{\mu} A_{\nu} A_{\lambda} 
- \hat{A}_{\mu} \partial_{\nu} \hat{A}_{\lambda} 
- \frac{2i}{3} \hat{A}_{\mu} \hat{A}_{\nu} \hat{A}_{\lambda} \right)\right. \nn\\  
\left. - \mathrm{Tr} D_{\mu} Y_A^{\dagger} D^{\mu} Y^A 
- i \mathrm{Tr} \; \psi^{A \dagger} \gamma^{\mu} D_{\mu} \psi_A
 - V_{\mathrm{bos}} - V_{\mathrm{ferm}} \right]
\ee
with the potentials
\be
V_{bos}= -\frac{4 \pi^2 }{3 k^2} \mathrm{Tr} \left( 
Y^A Y_A^\dagger Y^B Y_B^\dagger Y^C Y_C^\dagger
+ Y_A^\dagger Y^A Y_B^\dagger Y^B Y_C^\dagger Y^C \right. \nn\\
\left.  +4 Y^A Y_B^\dagger Y^C Y_A^\dagger Y^B Y_C^\dagger
-6 Y^A Y_B^\dagger Y^B Y_A^\dagger Y^C Y_C^\dagger \right),
\ee
and
\be
V_{ferm}= -\frac{2 i \pi }{k}
\mathrm{Tr} \left( Y_A^\dagger Y^A \psi^{B \dagger} \psi_B
-\psi^{B \dagger} Y^A Y_A^\dagger \psi_B 
-2 Y_A^\dagger Y^B \psi^{A \dagger} \psi_B
+2 \psi^{B \dagger} Y^A Y_B^\dagger \psi_A  \right. \nn\\
\left. -\epsilon^{ABCD} Y_A^\dagger \psi_B Y_C^\dagger \psi_D
+\epsilon_{ABCD} Y^A \psi^{B \dagger} Y^C \psi^{D \dagger} 
\right),
\ee
ABJM model actually has $SU(4)\sim SO(6)$ R-symmetry and $\cal N$=6  supersymmetry [15].
%%%%%%%%%%%%%%%%%%

As we want to consider Basu-Harvey type BPS equations, which have the dependence of only one of the spatial worldvolume coordinate, say $s$. The BPS equations of ABJM theory can be obtained by combining the kinetic terms and potential terms in the Hamiltonian and rewriting it as a sum of perfect squares plus some topological terms.  Denote $Y^A = (Z^1,Z^2,W^1,W^2)$ the formula used in this paper is [18]
\\
\be H &=&\int dx ds~{tr}(|\partial _{s}W^{\dagger A}+\frac{2\pi }{k}%
(W^{\dagger B}W_{B}W^{\dagger A}-W^{\dagger A}W_{B}W^{\dagger
B}-Z^{B}Z_{B}^{\dagger}W^{\dagger A}+W^{\dagger A}Z_{B}^{\dagger
}Z^{B})|^{2}  \nn \\
&&+|\partial _{s}Z^{A}+\frac{2\pi }{k}(Z^{B}Z_{B}^{\dagger
}Z^{A}-Z^{A}Z_{B}^{\dagger}Z^{B}-W^{\dagger
B}W_{B}Z^{A}+Z^{A}W_{B}W^{\dagger B})|^{2}  \nn \\
&&+\frac{16\pi ^{2}}{k^{2}}|\epsilon _{AC}\epsilon ^{BD}W_{B}Z^{C}W_{D}|^{2}+%
\frac{16\pi ^{2}}{k^{2}}|\epsilon ^{AC}\epsilon _{BD}Z^{B}W_{C}Z^{D}|^{2}) 
\nn \\
&&+\frac{\pi }{k}\int dx^{1}~{tr}(W_{A}W^{\dagger A}W_{B}W^{\dagger
B}-W^{\dagger A}W_{A}W^{\dagger B}W_{B}+2W^{\dagger
A}W_{A}Z^{B}Z_{B}^{\dagger}  \nn \\
&&-2W_{A}W^{\dagger A}Z_{B}^{\dagger}Z^{B}+Z_{A}^{\dagger
}Z^{A}Z_{B}^{\dagger}Z^{B}-Z^{A}Z_{A}^{\dagger}Z^{B}Z_{B}^{\dagger}),
\ee%
\\
in which the last term is topological and doesn't affect the dynamics in the bulk.  Therefore we have  a set of BPS equations, which minimize the energy in a given topological sector: 
\be
\partial _{s}W^{\dagger A}+\frac{2\pi }{k}(W^{\dagger B}W_{B}W^{\dagger
A}-W^{\dagger A}W_{B}W^{\dagger B}-Z^{B}Z_{B}^{\dagger}W^{\dagger
A}+W^{\dagger A}Z_{B}^{\dagger}Z^{B})=0
\ee
\be
\partial _{s}Z^{A}+\frac{2\pi }{k}(Z^{B}Z_{B}^{\dagger
}Z^{A}-Z^{A}Z_{B}^{\dagger}Z^{B}-W^{\dagger
B}W_{B}Z^{A}+Z^{A}W_{B}W^{\dagger B})=0  \label{Z_basu_01}
\ee
\be
\epsilon _{AC}\epsilon ^{BD}W_{B}Z^{C}W_{D}=\epsilon ^{AC}\epsilon
_{BD}Z^{B}W_{C}Z^{D}=0
\ee
The topological term gives the energy of the configuration when the BPS equations are satisfied.
\subsection{BPS solution and M2-M5  Systems}
To solve the BPS equations (2.5)-(2.7) one first  assumes $W^A=0$ and remained work is just to solve the following simple equations [16-18]:
\be 
\partial _{s}Z^{A}+\frac{2\pi }{k}(Z^{B}Z_{B}^{\dagger
}Z^{A}-Z^{A}Z_{B}^{\dagger}Z^{B})=0, \ee
where $A,B=1,2$.  Next, we separate the $s$-dependent and independent part by  
\be
Z^{A}=f(s)G^{A},~~~~~f(s)=\sqrt{\frac{k}{4\pi s}},\ee
where $G^{A}$s are $N\times N$matrices satisfying 
\be
G^{A}=G^{B}G_{B}^{\dagger}G^{A}-G^{A}G_{B}^{\dagger}G^{B}.
\ee
This equation is solved in [16,17] by first diagonalizing $G_{1}^{\dagger}~$ using the $ U(N)\times U(N)$ transformations and find that the another matrix $
G_{2}^{\dagger}~$ must be off-diagonal. The  $N$ dimensional irreducible solution is , 
\be
(G_{1}^{\dagger})_{m,n} =\sqrt{m-1}\delta _{m,n},~~~(G_{2}^{\dagger
})_{m,n}=\sqrt{N-m}\delta _{m+1,n}.\ee
The M5-brane tension ($T_5$) calculated from the above BPS funnel solution is consistent with the well known relation $2\pi T_5=T_2^2$ [22].  Above solution  was conjectured to be the only one in the BPS equations [17] and had also been use to study the vacuum of mass deformed ABJM [17,21].  

In this section we have collected the formula in [18] for self-contained.   In the next section we will find other possible solutions of BPS equation (2.5) -(2.7),  which can describe M2-KK6 (2D membrane, 6D Kaluza-Klein monopole) systems. 
%%%%%%%%%%%%%%%%%%%%%%%%%%%%
\section{New BPS Solutions : Numerical Searches}
We use the following matrix notation 
\be Y^A = (X,Z,Y,W) f(s)\ee
in which the funnel function $f(s)$ is described in (2.9).  We will first diagonalize $X$ by using the $ U(N)\times U(N)$ transformations and then search the possible non-trivial solutions for diagonalized $Y$ in the case of $W=0$ in this paper.  Once $X$ and $Y$ are known then $Z$ is more easily to be determined by BPS equations.  

  Let us first present our numerical searches for the solutions.  
\\
\\
1. The most simple solution of (2.11) is $2\times 2$ matrix
\be X=diag (0,1),~~~Y=diag(0,0),~~~~Z^{\dagger}=\left(\ba  {cc} 0&1\\0&0\ea \right)\ee
It represents funnel solution of 2M2-M5 system [16-18].
\\
\\
2. For the $3\times 3$ matrix the solution of (2.11) is  
\be X=diag (0,1,\sqrt2),~~~Y=0,~~~~Z^{\dagger}=\left(\ba  {ccc} 0&\sqrt2&0\\ 0&0&1\\0&0&0\ea \right)
\ee 
It represents funnel solution of 3M2-M5 system [16-18].   Our find is the following new solution
\be X=diag (0,0,1),~~~Y=diag(1,0,0),~~~~Z^{\dagger}=\left(\ba  {ccc} 0&1&\\ 0&{\bf 0}&1\\&0&0\ea \right)\ee
in which $Z_{3\times 3}$ looks, more or less,  like as the overlap of double $Z_{2\times 2}$ matrix in (3.2), with an overlapped element ${\bf 0}$.  Anyway, this is indeed a new irreducible solution.  Note that \{$X=diag (0,0,1)$, Y=diag(0,1,0)\} is an equivalent solution by just permuting matrix column and row elements.   However, neither \{$X=diag (0,0,1)$, Y=diag(0,0,1)\} nor \{$X=diag (1,0,1)$, Y=0\} is a  solution. 
\\
\\
3. For the $4\times 4$ matrix we could have the simple solution \{$X=diag (1,0,0,1), Y=diag(0,0,0,0)$\} which, however, is the reducible one formed by direct product of two  \{$X=diag (0,1)$, Y=0\} and is not a new solution. (Hereafter we neglect the reducible solutions.)  Of course, permutating X to become $X=diag (0,1,0,1)$ or  $X=diag (1,1,0,0)$ are also the possible solutions. (Hereafter we neglect these equivalent solutions.) The irreducible solution of $4\times 4$ matrix of (2.11) is  
\be X=diag (0,1,\sqrt 2,\sqrt 3),~~~Y=diag(0,0,0,0),~~~~Z^{\dagger}=\left(\ba  {cccc} 0&\sqrt 3&0&0\\0&0&\sqrt 2&0\\0&0&0&1\\0&0&0&0\ea \right)\ee
It represents funnel solution of 4M2-M5 system [16-18].  

For the $4\times 4$ matrix we have found the following new irreducible solution
\be X=diag (0,0,1,\sqrt 2),~~~Y=diag(1,0,0,0),~~~~Z^{\dagger}=\left(\ba  {cccc} 0&1&&\\0&\bf0&\sqrt 2&0\\&0&0&1\\&0&0&0\ea \right)\ee
in which $Z_{4\times 4}$ looks, more or less,  like as the overlap of $Z_{2\times 2}$ matrix in (3.2) with $Z_{3\times 3}$ matrix in (3.3), with overlapped element ${\bf 0}$.
With exchange $X\leftrightarrow Y$ we also have the equivalent solution, dues to the SU(4) symmetry in ABJM theory. (Hereafter we neglect these equivalent solutions.)
\\
\\
4. For the $5\times 5$ matrix, besides  \{$X=diag (0,1,\sqrt 2,\sqrt3,\sqrt4)$, Y=0\} [16-18] we have found the  two new irreducible solutions.  The first is
\be X=diag(0,0,1,\sqrt 2,\sqrt 3),~~~Y=diag(1,0,0,0,0),~~~~Z^{\dagger}=\left(\ba  {ccccc} 0&1&&&\\0&\bf0&\sqrt 3&0&0\\&0&0&\sqrt 2&0\\&0&0&0&1\\&0&0&0&0\ea \right)\ee
in which $Z_{5\times 5}$ looks, more or less,  like as the overlap of $Z_{2\times 2}$ matrix in (3.2) with $Z_{4\times 4}$ matrix in (3.5), with overlapped element ${\bf 0}$.   The second is 
\be X=diag(0,0,0,1,\sqrt 2),~~~Y=diag(\sqrt 2,1,0,0,0),~~~~Z^{\dagger}=\left(\ba  {ccccc} 0&1&0&&\\0&0&\sqrt 2&&\\0&0&\bf0&\sqrt 2&0\\&&0&0&1\\&&0&0&0\ea \right)\ee
in which $Z_{5\times 5}$ looks, more or less,  like as the overlap of double $Z_{3\times 3}$ matrix in (3.4), with an overlapped element ${\bf 0}$. (One of $Z_{3\times 3}$ has been permuted the order of colum and row.) 
\\
\\
5.  For the $6\times 6$ matrix, besides \{$X=diag (0,1,\sqrt 2,\sqrt3,\sqrt4,\sqrt5)$, Y=0\}  [16-18] we have found the following two  new non-equivalent irreducible solutions :
\be X=diag (0,0,1,\sqrt 2,\sqrt3,\sqrt4),~Y=diag(1,0,0,0,0,0),~Z^{\dagger}=\left(\ba  {cccccc} 0&1&&&&\\0&\bf0&\sqrt 4&0&0&0\\&0&0&\sqrt 3&0&0\\&0&0&0&\sqrt 2&0\\&0&0&0&0&1\\&0&0&0&0&0\ea \right)\\
\nn\\
X=diag (0,0,0,1,\sqrt 2,\sqrt3),~,Y=diag(\sqrt2,1,0,0,0,0),~Z^{\dagger}=\left(\ba  {cccccc} 0&1&0&&&\\0&0&\sqrt2&&&\\0&0&\bf 0&\sqrt 3&0&0\\&&0&0&\sqrt 2&0\\&&0&0&0&1\\&&0&0&0&0\ea \right)\nn
\\
\ee
As before, $Z_{6\times 6}$ looks, more or less,  like as the overlap of proper $Z$ matrix in (2.11), with an overlapped element ${\bf 0}$.  

Now we have seen a simple rule to find the general solution: Just overlap of two proper $Z$ matrix in (2.11), with an overlapped element ${\bf 0}$, and with the corresponding $X$ and $Y$ we then have a new solution. (Note that one of $Z$ matrix has permuted the order of column and row.)  {\bf The property of overlapping therein strongly suggests that this is a irreducible solution.} 
\\

 A further example is the case of $7\times 7$ matrix: Besides \{$X=diag (0,1,\sqrt 2,\sqrt3,\sqrt4,\sqrt5,\sqrt6)$, Y=0\}  [16-18] we have the three new non-equivalent irreducible solutions with non-zero matrix elements : 
\be X_{ii}=(0,0,1,\sqrt 2,\sqrt3,\sqrt4,\sqrt5),~Y_{ii}=(1,0,0,0,0,0,0),~Z_{i,i+1}^{\dagger}=(1,\sqrt5,\sqrt4,\sqrt3,\sqrt 2,1,0)\nn\\
\\
X_{ii}=(0,0,0,1,\sqrt 2,\sqrt3,\sqrt4),~Y_{ii}=(\sqrt2,1,0,0,0,0,0),~Z_{i,i+1}^{\dagger}=(1,\sqrt 2,\sqrt4,\sqrt3,\sqrt 2,1,0)\nn\\
\\
X_{ii}=(0,0,0,0,1,\sqrt 2,\sqrt3),~Y_{ii}=(\sqrt3,\sqrt2,1,0,0,0,0),~Z_{i,i+1}^{\dagger}=(1,\sqrt 2,\sqrt 3,\sqrt3,\sqrt 2,1)\nn\\
\ee
As before all above $Z_{7\times 7}$ look, more or less,  like as the overlap of proper $Z$ matrix in (2.11), with an overlapped element ${\bf 0}$. 

Besides the above solutions we also find other solutions
\be X=diag (c,0),~~~Y=diag(0,\sqrt {1-c^2}),~~~~Z^{\dagger}=\left(\ba  {cc} 0&1\\0&0\ea \right)\ee
\be X=diag (1,0),~~~Y=\left(\ba  {cc} 0&c\\0&0\ea \right),~~~~Z^{\dagger}=\left(\ba  {cc} 0&0\\\sqrt {1-c^2}&0\ea \right)\ee
for any $c$. However, we have not yet found a simple rule to generalize these solutions to higher dimensional matrix. 
%%%%%%%%%%%%%%%%%
\section{M2-KK6 :  BPS Solutions and KK6 Tension}
\subsection{General BPS Solutions in ABJM}
From above  study  we can write the new general solution of BPS equation with $W=0$.  

For $N\times N$ matrix  with $N=m+n+1$ we have general solution with following non-zero matrix elements:
\be
X_{i,i}&=&(0~~~,~~~0~~~~~~,...,~~0~,~0~,~0,~{\bf 0}~,1,\sqrt 2,\sqrt3,...,\sqrt {n-1},\sqrt n)\\
Y_{i,i}&=&(\sqrt m,\sqrt {m-1},...,\sqrt3,\sqrt2,~1,~{\bf 0}~,0,~~0,~~0,...,~~~~~0~~~,~~~0)\\
Z_{i,i+1}^{\dagger}&=&~~~(1,\sqrt 2,\sqrt3,...,\sqrt {m-1},\sqrt m,\sqrt n,\sqrt {n-1},...,\sqrt3,\sqrt 2,1)\ee
Notice that above solution is irreducible as Z matrix does not become a double block forms.   A very similar reducible solution constructed from (2.11) is the $N\times N$ matrix, $N=m+n+2$, with following non-zero matrix elements: 
\be
X_{i,i}&=&(0~~~,~~~0~~~~~~,...,~~0~,~0~,~0,~{\bf 0}~,~{\bf 0}~,1,\sqrt 2,\sqrt3,...,\sqrt {n-1},\sqrt n)\\
Y_{i,i}&=&(\sqrt m,\sqrt {m-1},...,\sqrt3,\sqrt2,~1,~{\bf 0}~,~{\bf 0}~, 0,~~0,~~0,...,~~~~~0~~~,~~~0)\\
Z_{i,i+1}^{\dagger}&=&~~~(1,\sqrt 2,\sqrt3,...,\sqrt {m-1},~{\bf 0}~,\sqrt m,\sqrt n,\sqrt {n-1},...,\sqrt3,\sqrt 2,1)\ee
This reducible solution had been had been used in [17,21] to count the vacua in the  mass deformed ABJM model.

  We now prove that our find matrix $X$, $Y$ and $Z$ in (4.1)-(4.3) is the  solution of BPS equations (2.5)-(2.7). We denote $X_{i,i}=x_i$, $Y_{i,i}=y_i$ and $Z_{i,i+1}=z_i$, then
\\
\\
$\bullet$ Step 1: As W=0 the first equation in (2.7) automatically be satisfied.
\\
\\
$\bullet$ Step 2 : The second equation in (2.7) becomes 
\be XWZ-ZWX =0 \Rightarrow x_i y_i z_i-z_iy_{i+1}x_{i+1}=0\ee
As $x_i y_i=0$ above equation is satisfied.
\\
\\
$\bullet$ Step 3 : Equation  (2.5) and equation  (2.6) become
\be  Y^{\dagger}-(-XX^{\dagger}Y^{\dagger}+Y^{\dagger}X^{\dagger}X+Y^{\dagger}Z^{\dagger}Z)=0 &\Rightarrow& y_i(-1+z_{i-1}^2-z_{i}^2)=0\\
 X-(ZZ^{\dagger}X-XZ^{\dagger}Z-Y^{\dagger}YX+XY^{\dagger}Y)=0 &\Rightarrow & x_i(-1+z_i^2-z_{i-1}^2)=0\\
 Z-(XX^{\dagger}Z-ZX^{\dagger}X-Y^{\dagger}YZ+ZY^{\dagger}Y)=0 &\Rightarrow & z_i(-1+x_i^2-x_{i+1}^2+y_{i+1}^2-y_i^2)=0\nn\\
\ee
It is easy to see that solutions of  (4.1)-(4.3) could satisfy above equation. Q.E.D.
\\

Using the above solution we can find  following trace value 
\be E_0(n)&\equiv & tr (XX^{\dagger}+YY^{\dagger}+ZZ^{\dagger}+WW^{\dagger})\nn\\
&=&n(n+1)+m(m+1)=N^2-N+2n^2-2n(N-1)
\ee
which will be proportional  to the energy associated with the BPS solutions, as studied in next subsection.   The minimum energy for a fixed $N$ is found to be at $n^*$ which is
\be \ba {ccc} n^*&={N-1\over2},~~~~~~&E_0^*={N^2-1\over2}, ~~~~~odd~N
\\
n^*&={N\over2},~~~~~~&E_0^*={N^2\over2}, ~~~~~~even~N\ea
\ee 
For the NM2-KK6 system we will analyze the lowest energy case among these solution.
%%%%%%%%%%%%%%%%%%%%
\subsection{Energy of BPS Solutions and KK6 Tension in ABJM}
The energy of the solution can be evaluated from  (2.4). 
\be E &=&\frac{\pi }{k}\int dx~\mbox{tr}(W_{A}W^{\dagger A}W_{B}W^{\dagger
B}-W^{\dagger A}W_{A}W^{\dagger B}W_{B}+2W^{\dagger
A}W_{A}Z^{B}Z_{B}^{\dagger}  \nn \\
&&-2W_{A}W^{\dagger A}Z_{B}^{\dagger}Z^{B}+Z_{A}^{\dagger
}Z^{A}Z_{B}^{\dagger}Z^{B}-Z^{A}Z_{A}^{\dagger}Z^{B}Z_{B}^{\dagger}) \nn\\
&=&2\int dsdx\mbox{tr}(\partial _{s}Z_{A}^{\dagger}\partial _{s}Z^{A}+\partial _{s}W_{A}^{\dagger}\partial _{s}W^{A})\nn\\
&=&{k\over 8\pi} E_0^*\int dsdx ~s^{-3}
\ee
in which $E_0^*$ is defined in (4.12).  We have used the BPS equations (2.5) and (2.6) to obtain the second line.  Case of $W=0$ in (4.13) reduces to that in [18]. 

After introducing the dimension parameter of M2 tension $T_{M2}$ [18] we can define the radius $R$ averaged over the M2 by
\be
R^2= {2{tr (Y^AY_A^{\dagger})}\over N}=\Big\{\ba {cc} {N^2-1\over N}{k\over4\pi T_{M2}}{1\over s},~~&odd~N \\
\\{N}{k\over4\pi T_{M2}}{1\over s},~~&even~N  \ea
\ee
In terms of $R$ the solution energy becomes
\be
E&=&{T_{M2}^2\over \pi}{N^2-1\over N^2} \int dx dR~{R^3\over k}~2\pi^2={T_{M2}^2\over \pi}{N^2-1\over N^2}\int dx~d^4y~{1\over k},~~odd~N \\
E&=&{T_{M2}^2\over \pi} \int dxdR~{R^3\over k}~2\pi^2 = {T_{M2}^2\over \pi}~\int dx~d^4y~{1\over k},~~~~~~~~~~~~~~~~~~~even~N 
\ee
Replacing ${N^2-1\over N^2}$ in the above equation by ${1\over2}{N\over N-1}$ produces  just the result in [18].  [18] explains that the factor $k$ in the denominator represents the fact that this M5-brane is divided by the $Z_{k}$ orbifold action, and $\frac{2\pi ^{2}}{k}~$is the volume of an $S^{3}/Z_{k}$.  As ${T_{M2}^2\over 2\pi}$ is equal to the M5 tension it is the funnel solution of $M2-M5$ system.  
\\

To proceed let us make following comments: 
\\
\\
1.  [18] considers only four non-zero coordinates, i.e. matrix $X$ and $Z$, and    interpretes its solution (2.11) as the M2-M5 system.  However, in our solution the values of $X,Y,Z$ are non-zero and we have six coordinates.  Thus the solution could not use to represent the M2-M5 systems. 
\\
\\
2.  The BPS equation in ABJM model has triplet matrix term, such as $W_B^{\dagger}W^BW_A^{\dagger}$. Use the radius defined in (4.14) this will always produce the $dR~R^3$ in the energy.  Thus, it always produces the volume integration $dx~d^4y$ which, at first sight, could not provide a sufficient space dimension, six, to represent the M2-KK6 system.
\\
\\
3.  Comment 1 says that our solution shall have 6-dimensional spaces while comment 2 say that we can have, at most, 5-dimensional spaces.  It seems to contradict to each other. The puzzle is solved in below.
\\

%%%%%%%%%%%%%%%%%%%%
 Let us first notice that  M2-brane can intersect with  KK6  over a 0-brane such that one of the worldvolume with coordinate ($t,x,z$) of the M-2-brane coincides with the isometry direction $(z)$ of the Taub-NUT space [8] (see (1.1)).   In this case M2 is a wrapped configuration [9].  Thus the integrations $\int dx$ in (4.15) and (4.16) shall be taken over a wrapped value of $\int dx= g_s\ell_s$.  (Note that $R_{11}=g_s\ell_s$) and (4.15) becomes
\\
\be
 E&=&{T_{M2}^2\over \pi}{N^2-1\over N^2}g_s\ell_s\int {d^4y\over k}={T_{M2}^2\over \pi}{N^2-1\over N^2}g_s\ell_s\cdot {1\over 4\pi \ell_s^2}\cdot\int {d^6y\over k}\nn\\
&=&{N^2-1\over N^2}T_{KK6}\int {d^6y\over k},~~~~~odd~N\\
\nn\\
E&=&{T_{M2}^2\over \pi}g_s\ell_s~\int {d^4y\over k}={T_{M2}^2\over \pi}g_s\ell_s\cdot {1\over 4\pi \ell_s^2}\cdot\int {d^6y\over k}\nn\\
&=&T_{KK6}\int {d^6y\over k},~~~~~even~N 
\ee
\\
in which we have added two sphere volume $\int dy^2=\int_{\ell_s} d\Omega_2=4\pi \ell_s^2 $. 

  This means that the KK6 in M2-KK6 system  described in ABJM theory shall be wrapped with 2 sphere with radium $\ell_s$.  Thus M2 membranes could expand into fuzzy $S^3$ plus a wrapped 2 sphere (with radium $\ell_2$) near the KK6 core.   In this interpretation we have a KK6 from M2 in ABJM theory. 

 It is amazing to see that the energy of M2-KK6 solution is independent of M2 number  in the case of even N.
%%%%%%%%%%%%%%
\section {Vacua of Mass Deformed ABJM and M2-2M5}
\subsection{Vacua of Mass Deformed ABJM}
In the mass deformed ABJM model [14,17,19] one adds the following mass term to Lagrangian (2.1)
\be
{\cal L}_{mass}= \mu^2Tr(Y^AY_A^{\dagger})
\ee
The vacuum is the solution with zero potential, which lead to the equations 
\be
0&=&\mu^2 W^{\dagger A}+\frac{2\pi }{k}(W^{\dagger B}W_{B}W^{\dagger
A}-W^{\dagger A}W_{B}W^{\dagger B}-Z^{B}Z_{B}^{\dagger}W^{\dagger
A}+W^{\dagger A}Z_{B}^{\dagger}Z^{B})\\
0&=&\mu^2 Z^{A}+\frac{2\pi }{k}(Z^{B}Z_{B}^{\dagger
}Z^{A}-Z^{A}Z_{B}^{\dagger}Z^{B}-W^{\dagger
B}W_{B}Z^{A}+Z^{A}W_{B}W^{\dagger B})
\ee
Denote $Z^A={1\over{\sqrt {2\pi\over \mu^2  k}}}(X,Z)$ and $W^A={1\over{\sqrt {2\pi\over \mu^2  k}}}(Y,0)$ it can be seen that (5.2) and (5.3) lead to (4.8)-(4.10).  Thus the vacua solutions of (5.2)-(5.3) are those found in (4.1)-(4.3).

  In the previous literature [17,21], the solution for M2-M5 system described in (2.11) is used to form the reducible matrix to count the vacua state characterized by M5-branes wrapping fuzzy 3-spheres.  In this case there is the only one possible solution for a fixed $N$.   However, Our solution found in  (4.1)-(4.3),  has many non-equivalent solutions for a fixed $N$.   Thus, the vacua characterized by KK6-branes wrapping fuzzy 3-spheres with wrapped 2 sphere (with radius $\ell_s$) are more complex then that in M2-M5 system. 
%%%%%%%%%%%%%%%
\subsection{M2-KK6 and M2-2M5}
It is interesting to mention that the authors in  [23] have found that the funnel solution (2.11) matches the D4-brane interpretation just as well as the M5-brane interpretation, after a properly wrapped about a suitable radius.  Motivated by this, although our solution is regarded as M2-KK6 systems we would try to give it following two interpretations.
\\

1.  We see that the tension of new BPS solution calculated in (4.15) is just the double value of M2-M5 system [18] in large $N$ limit.  Thus our solution may be used to represent the M2-M5-M5 system.
\be
 \left(0|M2,M5,M5\right)={\Big\{}\ba{lrcrrrrrrrr}
t  & x & z&-&-&-&-&-&-&-&-  \\
t  & x & -&x_1&x_2&z_1&z_2&-&-&- &-\\
t  & x & -&x_1&x_2&-&-&y_1&y_2&- &-
\ea
\ee
\\
The integration in energy formula (4.15)-(4.16) is now regarded as an integration by integrating over  space $(x,x_1,x_2,z_1,z_2)$ plus another one by integrating over space $(x,x_1,x_2,y_1,y_2)$.    Thus we could interprete our solution as the system of  two five-branes intersecting on a three-brane.  Note that M2-M5-M5 system is the $n=1/4$ supersymmetric system, as analyzed in [24]. 
\\

2. To begin with, let us mention following two issues. 

  First : Authors in  [23] argue that, as $G_1^{\dagger}=G_1$ the solution in (2.11) will represent the funnel solution with coordinates $G_1$, $G_2$ and $G_2^{\dagger}$, which in fact has only 3 coordinates and the funnel solution is fuzzy $S^2$ instead of $S^3$ regarded in [16-18]. ( Authors in [16-18] regard  $G_1$, $G_2$  as two complex matrix and thus four coordinates.)  

   Second :  For finite N, the field theory description [20]  shows that the ground states are labeled simply by partitions of N, rather than labeled partitions in [17]. 
[17] count vacua by using reducible matrix, which is clearly more numerous than partitions of N, so there is a discrepancy between the counting of classical vacua and the expected number of vacua for the theory. 

Given the subtleties related to the counting of vacua in the mass-deformed ABJM theory and diagonalilty of matric $X$ and $Y$ in (4.1) and (4.2) we have following another interpretations for our solutions.  {\bf Our solution has only 4 coordinates and it represents M2-2M5 system.}  This consists with the coordinate number and solution energy.  In this case,  the vacua are counted form multiple  number of the irreducible matrix, which are more complex then before.  The details remains to be investigated. 
\\

In conclusion, the BPS solution found in [16,17] is described M2-M5 system and, after properly being wrapped, it can describe D4 system [23].  In the similar way, our BPS solution in (4.1)-(4.3) may be used to describe M2-2M5 system and, after properly wrapped, it can describe M2-KK6 system.

%%%%%%%%%%%%%%%%%%%%%%
\section {Discussion}
In this paper we  first present our numerical  searches for BPS solutions in low dimensional matrix to find the M2-KK6 systems in ABJM theory.  From the numerical result we find a general rule to describe the general solution and prove that it satisfies the BPS equations in ABJM theory.   We calculate the tension of the solution and show a possible way to explain our solution as  describing the  M2-KK6 (2D membranes with 6D Kaluza-Klein monopole) systems.  We mention that our solutions also reveal new vacua in the mass deformed ABJM theory and, contrast to the previous investigation in which the vacua are reducible, ours may be irreducible. We also discuss a possible way to interprete  our new BPS solutions as a M2-2M5 system. 
\\

 Finally, let us mention following comments to conclude the paper.

1.    It is interesting to study the fluctuations of the ground-state/funnel solutions, along  the line of [23], to see whether these interpretations are correct.  Note that the study the fluctuations over the our new solution may produce the 2M5 action, which is the simplest case of multiple M5 [25], and is very interesting at present.  

2.  As our solution can describes M2-KK6 system with proper wrapped, and on other hand, it can also describes M2-2M5 system, it seems that wrapped KK6 is the configuration of 2M5.   Thus, the n wrapped KK6, n-KK6, may be the configuration of 2n-M5.  Along this line we can study the multiple M5 system.

3.  In [26] the BPS conditions preserving $n/12~(n = 1,\cdot\cdot\cdot,6)$ supersymmetries in the ABJM  model are investigated and the BPS equations are clearly classified.  It is interesting to see whether the other BPS objects therein could be derived form the BPS equations (2.5)-(2.7).   

The investigations of these problems are on progress.
\\
\\
%%%%%%%%%%%%%
{\bf Acknowledgments} :  This work is supported in part by the Taiwan National Science Council. 
%%%%%%%%%%%%%%%%%%%%%%
\\
\\
\begin{center} {\bf REFERENCES}\end{center}
%%%%%%%%%%%%%%%%%%%%%%
\begin{enumerate}
\item J. Bagger and N. Lambert, ``Modeling multiple M2s", Phys. Rev. D 75 (2007) 045020  [arXiv:hep-th/0611108]; J. Bagger and N. Lambert,``Gauge Symmetry and Supersymmetry of Multiple M2-Branes", Phys. Rev. D 77 (2008) 065008  [arXiv:0711.0955 [hep-th]]; J. Bagger and N. Lambert, ``Comments On Multiple M2-branes", JHEP 0802 (2008) 105  [arXiv:0712.3738 [hep-th]].
\item A. Gustavsson, ``Algebraic structures on parallel M2-branes", Nucl.Phys.B811 (2009) 66  [arXiv:0709.1260 [hep-th]]. 
\item V. T. Filippov, ``n-Lie algebras", Sib. Mat. Zh.,26, No. 6 (1985) 126140;\\  Jose A. de Azcarraga, Jose M. Izquierdo ``n-ary algebras: a review with applications" [arXiv:1005.1028 [math-ph]].
\item M. Van Raamsdonk, ``Comments on the Bagger-Lambert theory and multiple M2-branes", JHEP0805 (2008) 105 [arXiv:0803.3803 [hep-th]]; \\N. Lambert and D. Tong, ``Membranes on an Orbifold", Phys.Rev.Lett.101 2008 041602 [arXiv:0804.1114 [hep-th]];\\ J. Distler, S. Mukhi, C. Papageorgakis and M. Van Raamsdonk, ``M2-branes on M-folds", JHEP 0805 (2008) 038 [arXiv:0804.1256 [hep-th].
\item Y. Nambu, ``Generalized Hamiltonian dynamics", Phys. Rev. D 7 (1973) 2405; L. Takhtajan, ``On Foundation Of The Generalized Nambu Mechanics (Second Version)", Commun. Math. Phys. 160 (1994) 295  [arXiv:hep-th/9301111].
\item  D. Alekseevsky, P. Guha, ``On Decomposability of Nambu-Poisson Tensor", Acta. Math. Univ. Commenianae 65 (1996) 1.
\item P.-M. Ho, R.-C. Hou, and Y. Matsuo, ``Lie 3-algebra and multiple M2-branes", JHEP 06 (2008) 020, [arXiv:0804.2110 [hep-th]]; P.-M. Ho and Y. Matsuo, ``M5 from M2", JHEP 06 (2008) 105, [arXiv:0804.3629 [hep-th]];\\ P.-M. Ho, Y. Imamura, Y. Matsuo, and S. Shiba, ``M5-brane in three-form flux and multiple M2-branes", JHEP 08 (2008) 014, [arXiv:0805.2898 [hep-th]].
\item  R.D. Sorkin, Phys. Rev. Lett. 51 (1983) 87;\\ D.J. Gross and M. Perry, Nucl. Phys. B226 (1983) 29.
\item  E. Bergshoeff, M. de Roo, E. Eyras, B. Janssen and J.P. van der Schaar, ``Intersections involving monopoles and waves in eleven dimensions", Class. Quantum Grav. 14 (1997) 2757, [hep-th/9704120];\\ E. Bergshoeff,  E. Eyras,  and Y. Lozano, ``The Massive Kaluza-Klein Monopole", Phys.Lett. B430 (1998) 77-86 [hep-th/9802199].
\item E. Bergshoeff, B. Janssen and T. Ortin, ``Kaluza-Klein Monopoles and Gauged Sigma-Models", Phys. Lett. B410 (1997) 132 [hep-th/9706117].
\item C.M. Hull, ``Gravitational Duality, Branes and Charges", Nucl. Phys. B509 (1998) 216, [hep-th/9705162];\\  E. Bergshoeff, J. Gomis, P. K. Townsend, ``M-brane intersections from worldvolume superalgebras," Phys.Lett. B421 (1998) 109 [hep-th/9711043].
\item  Wung-Hong Huang, ``KK6 from M2 in BLG " JHEP 1009 (2010) 109 [arXiv:1006.4100 [hep-th]] 
\item O. Aharony, O. Bergman, D. L. Jafferis and J. Maldacena, ``N=6 superconformal Chern-Simons matter theories, M2-branes and their gravity duals," JHEP 0810 (2008) 091  [arXiv: 0806.1218 [hep-th]].
\item O. Aharony, O. Bergman and D. L. Jafferis, ``Fractional M2-branes,"
JHEP 0811(2008)  043  [arXiv:0807.4924 [hep-th]]; [9] K. Hosomichi, K. M. Lee, S. Lee, S. Lee and J. Park, ``N=5,6 Superconformal Chern-Simons Theories and M2-branes on Orbifolds," JHEP 0809 (2008)  002  [arXiv:0806.4977 [hep-th]].
\item  M. Benna, I. Klebanov, T. Klose and M. Smedback, ``Superconformal Chern-Simons Theories and AdS(4)/CFT(3) Correspondence,"  JHEP 0809 (2008) 072 (2008) [arXiv:0806.1519].
\item S. Terashima, ``On M5-branes in N = 6 membrane action", JHEP 0808 (2008) 080 [arXiv:0807.0197[hep-th]].
\item  J. Gomis, D. Rodriguez-Gomez, M. Van Raamsdonk, and H. Verlinde, ``A Massive Study of M2-brane Proposals," JHEP 09 (2008) 113 [arXiv:0807.1074 [hep-th]].
\item K. Hanaki and H. Lin H, ``M2-M5 systems in N = 6 Chern Simons theory", JHEP 0909  (2008)  067 [arXiv:0807.2074[hep-th]].
\item K. Hosomichi, K. M. Lee, S. Lee, S. Lee and J. Park, ``N=4 Superconformal Chern-Simons Theories with Hyper and Twisted Hyper Multiplets," JHEP 0809 (2008) 072 [arXiv:0805.3662 [hepth]]; K. Hosomichi, K. M. Lee, S. Lee, S. Lee and J. Park, ``N=5,6 Superconformal Chern-Simons Theories and M2-branes on Orbifolds,"  JHEP0809 (2008) 002 [ arXiv:0806.4977 [hep-th]].
\item H. Lin, O. Lunin and J. M. Maldacena, ``Bubbling AdS space and 1/2 BPS geometries,"  JHEP 0410 (2004) 025 [arXiv:hep-th/0409174]; D. E. Berenstein, J. M. Maldacena and H. S. Nastase, ``Strings in flat space and pp waves from N = 4 super Yang Mills," JHEP 0204 (2002) 013 (2002) [arXiv:hep-th/0202021];  K. Dasgupta, M. M. Sheikh-Jabbari and M. Van Raamsdonk, ``Matrix perturbation
theory forM-theory on a PP-wave," JHEP 0205 (2002) 056 (2002) [arXiv:hep-th/0205185].
\item  H.-C. Kim and S. Kim, ``Supersymmetric vacua of mass-deformed M2-brane theory,"  Nucl. Phys. B839 (2010) 96 [arXiv:1001.3153[hep-th]].
\item J.~H.~Schwarz, ``The power of M theory,'' Phys. Lett. B 367 (1996) 97  [arXiv:hep-th/9510086];  M.~J.~Duff, J.~T.~Liu and R.~Minasian,
``Eleven-dimensional origin of string / string duality: A one-loop test,''
Nucl. Phys. B261 (1995) 261 [arXiv:hep-th/9506126]. 
\item H. Nastase, C. Papageorgakis, and S. Ramgoolam, ``The fuzzy S2 structure of M2-M5 systems in ABJM membrane theories," JHEP 05 (2009) 123 [ arXiv:0903.3966 [hep-th]]
\item D. S. Berman and N. B. Copland, ``Five-brane Calibrations and Fuzzy Funnels," Nucl.Phys. B723 (2005) 117 [arXiv:hep-th/0504044].
\item N. Lambert and C. Papageorgakis, ``Nonabelian (2,0) Tensor Multiplets and 3-algebras," JHEP 1008 (2010) 083 [arXiv:hep-th1007.2982]; Kuo-Wei Huang and Wung-Hong Huang, ``Lie 3-Algebra Non-Abelian (2,0) Theory in Loop Space," arXiv:1008.3834. 
\item T. Fujimori, K. Iwasaki, Y. Kobayashi, S. Sasaki, ``Classification of BPS Objects in N=6 Chern-Simons Matter Theory,"  JHEP 1010 (2010) 002  [arXiv:1007.1588]

\end{enumerate}
\end{document}